\documentstyle[12pt,epsfig]{article}

\begin{document}

\begin{titlepage}
\vskip1.5cm
\begin{center}

{\Large \bf Binding energy corrections in positronium decays}
\vskip 2cm

{\large G. L\'opez Castro$^1$, J. Pestieau$^2$, and C. Smith$^2$} \\

\

$^1$ {\it Departamento  F\'\i sica, Centro de Investigaci\'on y de
Estudios} \\ {\it Avanzados del IPN, Apdo. Postal 14-740, 07000 M\'exico,
D.F., M\'exico} \\

$^2$ {\it Institut de Physique Th\'eorique, Universit\'e catholique de \\
Louvain, Chemin du Cyclotron 2, B-1348 Louvain-la-Neuve, Belgium} 
 \end{center}
\vskip1.5cm
\begin{abstract}
Positronium annihilation amplitudes that are computed by  assuming a
factorization approximation with on-shell intermediate leptons, do not
exhibit good analytical behavior. We propose an {\it ansatz} which
allows to include binding energy corrections and obtain the correct
analytical and gauge invariance behavior of these QED amplitudes. As a
consequence of these non-perturbative corrections, the parapositronium and
orthopositronium decay rates receive corrections of order $\alpha^4$ and
$\alpha^2$, respectively. These new corrections for orthopositronium are
relevant in view of a precise comparison between recent theoretical and
experimental developments. Implications are pointed out for analogous
decays of quarkonia .

 \end{abstract}


\end{titlepage}%

\medskip

\

Parapositronium (p-Ps) and orthopositronium (o-Ps) are bound states of
$e^+e^-$ whose lifetimes for 2$\gamma$ and 3$\gamma$ channels,
respectively, have been measured with high precision:
\begin{eqnarray}
\Gamma(\mbox{\rm p-Ps} \rightarrow \gamma\gamma) &=& 7090.9(1.7)\ \mu
s^{-1}
\cite{exp1}
\\ 
\Gamma(\mbox{\rm o-Ps} \rightarrow \gamma\gamma\gamma) &=& \left\{
\begin{array}{c} 
7.0514(14)\ \mu s^{-1} \cite{exp2}, \\
7.0482(16)\  \mu s^{-1} \cite{exp3}, \\
7.0398(29)\  \mu s^{-1} \cite{exp4} \ .
\end{array} \right.
\end{eqnarray}
The corresponding theoretical predictions which include  perturbative
QED corrections to a non-relativistic treatment of the bound state wave
function, have been computed also with high accuracy
\cite{radcor,adkins}:
\begin{eqnarray}
\Gamma(\mbox{\rm p-Ps} \rightarrow \gamma\gamma) &=& \frac{\alpha^5 m}{2}
\left(
1-(5-\frac{\pi^2}{4}) \frac{\alpha}{\pi} + 2\alpha^2 \ln \frac{1}{\alpha}
+ 1.75(30) \left(\frac{\alpha}{\pi} \right)^2 - \frac{3\alpha^3}{2\pi}
\ln^2 \frac{1}{\alpha} \right) \nonumber \\
&=& 7989.50(2)\ \mu s^{-1} \\
\Gamma(\mbox{\rm o-Ps} \rightarrow \gamma\gamma\gamma) &=&
\frac{2(\pi^2-9)\alpha^6
m}{9\pi} \left( 1 -10.28661(1) \frac{\alpha}{\pi} -\frac{\alpha^2}{3} \ln
\frac{1}{\alpha} + B_0 \left( \frac{\alpha}{\pi} \right)^2 -
\frac{3\alpha^3}{2\pi} \ln^2 \frac{1}{\alpha} \right) \nonumber \\
&=& (7.0382 + B_0\: 0.39 \times 10^{-4})\ \mu s^{-1} \ ,
\end{eqnarray}
where $m$ denotes de mass of the electron and $\alpha$ is the fine
structure constant. As it can be observed, predictions for parapositronium
are in very good agreement with experiment, while the experimental results
for orthopositronium reported in Refs. \cite{exp2,exp3} largely disagree
with the theoretical expectation.
  Recent theoretical efforts  have focused on a more complete evaluation
of the non-logarithmic $O(\alpha^2)$ perturbative corrections for
orthopositronium, with the result $B_0=44.52(26)$ \cite{adkins1} or,
equivalently,
\begin{equation}
\Gamma(\mbox{\rm o-Ps} \rightarrow \gamma\gamma\gamma)= 7.039934(10)\ \mu
s^{-1} \ .
\end{equation}
This result renders even closer the theoretical prediction to the
experimental measurement of Ref. \cite{exp4}.

 Although these achievements of perturbative QED for bound states are
impressive, little is known 
about the effects of non-perturbative corrections. In the present paper we
implement a mechanism, imposed by analyticity and gauge invariance, which
allows to introduce binding energy (BE) effects in
the positronium decay amplitudes. These new corrections affect the $\gamma
\gamma$ decay rate of parapositronium at the order $\alpha^4$, while
corrections of order $\alpha^2$ are induced in the orthopositronium decay
rate and in the single photon spectrum of p-Dm$\rightarrow 
e^+e^-\gamma$ and o-Ps$\rightarrow \gamma \gamma \gamma$ decays (p-Dm
(o-Dm) denotes the J=0(1) bound state of the dimuonium system
$\mu^+\mu^-$).
Moreover, the single photon energy spectrum in the last two decays
previously mentioned is softened with respect to results obtained when
BE effects are neglected. Our
implementation of these non-perturbative corrections relies
on well grounded basis of QED such as the requirement of a correct
behavior of the decay amplitude in terms of the photon energies
(analyticity) and on electromagnetic gauge invariance. We briefly discuss
at the end, the possible implications for analogous decays of quarkonium
states. Details of our calculations will be given elsewhere \cite{next}.

   Let us first try to explain how problems related to analyticity of
the positronium decay amplitudes appears in the usual approaches. Current
calculations of the
positronium rates assume a factorization approximation of the
dynamics contained in the decay amplitude. The amplitudes for positronium
decays into a given final state $X$ is approximated as the product 
of the amplitude for the bound state annihilation (described by
$\Phi_0$, the $e^+e^-$ wavefunction at the origin), times the annihilation
amplitude of free leptons into the final state $X$. Under this assumption,
the positronium decay rates can be written as \cite{wheeler}:
\begin{equation}
\Gamma(\mbox{\rm Ps}(^{2J+1}S) \rightarrow X) = \frac{1}{2J+1}|\Phi_0|^2
\cdot (v_{rel}
\sigma(e^+e^-
\rightarrow X))_{v_{rel} \rightarrow 0}\ ,
\end{equation}
where $ \sigma(e^+e^-\rightarrow X)$ denotes the total cross section for
$e^+e^- \rightarrow X$ and $v_{rel}$ is the relative velocity of the
constituents in their center of mass frame.

  Since the leptons in the intermediate stage are taken on their
mass-shell, the annihilation amplitude for $e^+e^- \rightarrow n \gamma$
is automatically gauge-invariant. However, in this approximation the
positronium decay amplitudes do not exhibit the expected analytical
behavior. Indeed, for definiteness let us consider  the
3$\gamma$ decay mode of orthopositronium where the photon energy is
kinematically allowed to vanish. The behavior of the $e^+e^-
\rightarrow 3\gamma$ amplitude in terms
of the photon momenta  is driven in the soft-photon limit by the
intermediate electron
propagators as follows:
\begin{equation}
\frac{i}{\not k- \not l_i - m} = \frac{i(\not k- \not l_i+
m)}{-2l_i
\cdot k}\ ,
\end{equation}
where $k$ is the four-momentum of the electron that emits a photon of
four-momentum $l_i$. Note that in the soft-photon limit
$l_i \rightarrow 0$, the positronium decay amplitude seems to diverge as 
$l_i^{-1}$ . Actually, this is only apparent since selection rules
cancels these infrared divergencies in the static limit
($k=(m,0,0,0))$, and the amplitude starts indeed at order $l_i^0$. This is
in contradiction with the fact that in the soft-photon  limit the
amplitude must vanish \cite{low} since o-Ps $\rightarrow \gamma \gamma
\gamma$ involve only neutral external bosons.

   We can try to cure this bad analytical behavior by realizing that
electrons in the intermediate stage are always off their mass-shell due
to BE effects. Actually, the $e^+e^-$ bound states involve two
mass scales: the mass $M$ of the bound state and the mass $m$ of the
constituent electrons. These masses differ by terms of order $\alpha^2$
and are related through the binding energy $E_{bind}$,
\begin{equation}
E_{bind} \equiv M-2m = -\frac{1}{4} m\alpha^2\ .
\end{equation}
Thus, the relevant momentum scale for intermediate electrons is determined
by $M/2$. In this case, the lepton propagator involved in the $e^+e^-
\rightarrow 3\gamma$ decay amplitude becomes ($k^2=M^2/4$):
\begin{equation}
\frac{i}{\not k-\not l_i- m} = \frac{i(\not k-\not l_i + m)}{-2l_i 
\cdot k-\gamma^2}\ ,
\end{equation}
where $\gamma^2 \equiv m^2-M^2/4 \approx M^2 \alpha^2/16$, takes into
account the bound state nature of the $e^+e^-$ pair in the rescattering
process. The presence of $\gamma^2$ in the denominator would provide to
the decay amplitude a better analytical behavior. Unfortunately, the
gauge invariance of the $e^+e^- \rightarrow \gamma\gamma\gamma$ amplitude
is spoiled due to the presence of the two mass scales $M$ and $m$. Thus, a
more sophisticated procedure is required to restore analyticity without
destroying gauge invariance.

   A natural way to incorporate BE effects is by considering
a model with a loop of virtual leptons for the decays of positronium at
lowest order. For
definiteness we consider the $n\gamma$ decay of the positronium state
$B(^{2J+1}S)$: $J=0 (1)$ being the spin of the p-Ps(p-Dm) state, $n=2$ for
p-Ps or p-Dm, and $n=3$ for o-Ps or o-Dm decays (the $e^+e^-\gamma$ mode
of paradimuonium can be reached from p-Dm$ \rightarrow \gamma \gamma^*$
where $\gamma^*$ is a virtual photon). In this model we need to introduce
a coupling 
$F_B\Gamma$ to describe the $Be^+e^-$ vertex, where $\Gamma=\gamma_5
(\not \eta)$ for para(ortho)-positronium decay, $\eta_{\alpha}$ is the
polarization four-vector of the $B(^3S)$ state and $F_B$ is a form factor
that describes the structure of the vertex. The evaluation of the
$B(^{2J+1}S) \rightarrow X$ decay amplitude follows standard rules (see
Fig. 1).

  Applying the Feynman rules to Fig. 1 the decay amplitude is given by
(contraction with photon polarization vectors $\epsilon_{\mu}(l_1)
\cdots$ must be understood):
 \begin{eqnarray}
{\cal M}^{\mu\nu\cdots}(B \rightarrow n\gamma) &=& \int \frac{d^4
q}{(2\pi)^4}
F_B {\rm Tr} \left\{ \Gamma \frac{i(\not q-\frac{\not P}{2}+m)}{
(q-\frac{P}{2})^2-m^2} \Gamma^{\mu \nu \cdots} \frac{i(\not q+\frac{\not
P}{2}+m)}{ (q+\frac{P}{2})^2-m^2} \right\}\ ,
\end{eqnarray}
where $\Gamma^{\mu \nu \cdots}$ is a properly symmetrized amplitude for
annihilation of virtual lepton pairs into $n \gamma$. It is important to
emphasize that this amplitude is gauge-invariant if $F_B$ is constant.

  To clearly illustrate how this loop model reproduces, in the limit of
zero BE,  the lowest order amplitudes of the on-shell 
approximation, let us consider the following {\it ansatz} for $F_B$:
\begin{equation}
F_B = i C\Phi_0  \frac{8\pi \gamma}{({\bf q}^2
+\gamma^2)^2} \cdot ({\bf q}^2 +\gamma^2)
\end{equation}
where $\gamma$ contains the BE, 
$\Phi_0=\sqrt{\alpha^3m^3/8\pi}$ is the ground state wavefunction 
of $e^+e^-$ at the origin, and $C=-2/\sqrt{M}$ is a
normalization constant.

  Using well known representations of Dirac-delta functions \cite{delta}  
we can check that in the limit of zero BE ($\gamma \rightarrow 0,\ m-M/2
\rightarrow 0$) we obtain:
\begin{equation}
\frac{F_B}{(q-\frac{P}{2})^2-m^2} \cdot \frac{1}{(q+\frac{P}{2})^2-m^2}
\rightarrow \frac{1}{2}C\Phi_0 (2\pi)^4\delta^{(4)}(q)
\frac{\sqrt{{\bf q}^2+m^2}+\frac{M}{2}}{q_0^2-(\sqrt{{\bf
q}^2+m^2}+\frac{M}{2})^2} \ .
\end{equation}

Thus, upon (trivial) integration of Eq. (10) one gets:
\begin{equation}
{\cal M}^{\mu \nu \cdots}(B \rightarrow n\gamma)=
\frac{C\Phi_0}{8M} {\rm Tr}\left\{ \Gamma \left(\not
P-M \right) \overline{\Gamma}^{\mu \nu \cdots} \left(\not
P+M  \right) \right\} \ ,
\end{equation}
where $\overline{\Gamma}^{\mu \nu \cdots}$ denotes the {\it reduced} 
vertex evaluated at $q=0$. Note that the factors $(\not P \pm M)$
play the role of projector operators external to the action of photon
vertices.

  Setting in the rest frame of positronium and using $2m=M$ (and the
condition $P_{\alpha}\eta^{\alpha}=0$ for orthopositronium) we arrive at
the well known results \cite{adkins} of the factorization approximation,
namely:
\begin{equation}
{\cal M}^{\mu \nu \cdots}(B \rightarrow n\gamma)=
-C\Phi_0\frac{M}{2\sqrt{2}} {\rm Tr}\left\{
\frac{1+\gamma_0}{\sqrt{2}}\Gamma 
\overline{\Gamma}^{\mu \nu \cdots}  \right\} \ .
\end{equation}

  Let us return to the issues concerning gauge-invariance in the context
of the present model when BE is {\it not} neglected. Gauge
invariance requires that the amplitude in Eq.
(10) (contracted with photon polarizations) vanishes when
$\epsilon_{\alpha}(l_i) \rightarrow l_{i\alpha}$ for
any external photon. The amplitude for p-Ps$\rightarrow \gamma \gamma$ is
always gauge-invariant, no matter the specific form of $F_B$. This
follows from the fact that Lorentz covariance for the
axial-$\gamma\gamma$ vertex implies that the amplitude should corresponds
to an effective operator $\sim F_{\mu\nu}\widetilde{F}^{\mu\nu}$, which is
automatically gauge-invariant. This is not the case for o-Ps$
\rightarrow \gamma\gamma \gamma$ decays. In this case the amplitude of
Eq. (10) is gauge invariant provided $F_B$ remains the same under the
shifts $q \rightarrow q+l_i$ of the integration variable for all photons.
Since this is achieved only if $F_B$ is constant, it means
that other contributions should be added to the
orthopositronium decay amplitude in this loop model in order to
compensate for this lack of gauge invariance. 

In view of these difficulties, we propose and {\it ansatz} for positronium
decay amplitudes that fulfills gauge invariance and analyticity
simultaneously. Our recipe contains  three steps: $(a)$ evaluate the
expression $\overline{\Gamma}^{\mu\nu ...}$ in Eq. (13)  for constituents
masses $M/2$ (this modified expression will be denoted by
$\overline{\Gamma}^{\mu\nu ...}_{M/2}$); this will ensure gauge
invariance, $(b)$ introduce BE effects by multiplying the
amplitude $\overline{\cal M}(B(^{2J+1}S) \rightarrow n\gamma)$ ({\it
i.e.} the amplitude of Eq. (13) where we  replace $m$ by $M/2$ in the
argument of the Trace operator) by a factor:
${\cal A}_1$ for p-Ps (p-Dm) states and $\prod_{i=1}^3 {\cal A}_i)$ for
o-Ps (o-Dm), where we have defined ${\cal A}_i\equiv
P.l_i/(P.l_i+\gamma^2)$ (note that ${\cal A}_1={\cal A}_2$ for
p-Ps$\rightarrow 2\gamma$). This will warrant the
correct analytical properties of the amplitude in the soft-photon limit
and, $(c)$ obtain the decay rate by integration over the physical
phase-space determined by the mass $M$.

   Let us first note that the proposed ansatz is fulfilled automatically
for the p-Ps$ \rightarrow \gamma\gamma$ decay. The reduced vertex in Eq.
(13) is given by (remember $q=0$ and $P^2=M^2$):
\begin{eqnarray}
\overline{\Gamma}^{\mu\nu}&=&(-ie)^2 \frac{\gamma^{\mu} \left(
-\frac{\not
P}{2} + \not l_1+m \right) \gamma^{\nu}+ \gamma^{\nu} \left( \frac{\not
P}{2} - \not l_1+m \right) \gamma^{\mu}}{\left( \frac{P}{2}-l_1
\right)^2-m^2} \nonumber \\
&=& (-ie)^2 \frac{\gamma^{\mu} \left( -\frac{\not P}{2} 
+ \not l_1+m \right) \gamma^{\nu}+ \gamma^{\nu} \left(
\frac{\not P}{2} - \not l_1+m \right) \gamma^{\mu}}{\left( \frac{P}{2}-l_1
\right)^2-\frac{M^2}{4}} \left( {\cal A}_1 \right) \ .
\end{eqnarray}

In the second line of Eq. (15) we can replace  $m\rightarrow M/2$ in
the numerator because terms proportional to $m$ in
$\overline{\Gamma}^{\mu\nu}$ cancel when performing the trace in Eq.
(13). Therefore, gauge invariance is preserved independently of $m$ and we
have:
\begin{equation}
{\cal M}^{\mu \nu}(\mbox{\rm p-Ps} \rightarrow \gamma\gamma) =
\overline{\cal M}^{\mu \nu}(\mbox{\rm
p-Ps} \rightarrow \gamma\gamma) 
=\frac{C\Phi_0}{8M} {\rm Tr} \left\{
\gamma_5(\not P-M) \overline{\Gamma}^{\mu\nu}_{M/2} (\not P+M) \right\}
{\cal A}_1 \ .
\end{equation}
This amplitude satisfies analyticity, as required.

  Secondly, the reduced vertex for orthopositronium decay
$\overline{\Gamma}^{\mu\nu\rho}$ can be worked in the following way. In
order to accomplish gauge invariance, we are forced to replace $m$ by
$M/2$
for the mass of the constituents and simultaneously add a third analytical
factor $A_i$ \cite{next} to each of the six amplitudes contributing to
$\overline{\Gamma}^{\mu\nu\rho}$. This gives rise to the final
gauge-invariant and analytical amplitude for orthopositronium decay:
\begin{equation}
\overline{\cal M}^{\mu \nu\rho} =\frac{C\Phi_0}{8M} {\rm Tr} \left\{
\not \eta (\not P-M) \overline{\Gamma}^{\mu\nu\rho}_{M/2} (\not P+M)
\right\} \prod_{i=1}^3{\cal A}_i \ ,
\end{equation}
where $\eta_{\alpha}$ represents the polarization four-vector of 
orthopositronium. In the soft photon limit ($l_i \rightarrow 0$),
$\overline{\cal M}^{\mu \nu \rho}$ vanishes as required.

  Up to now we have considered the effects of nonzero BE corrections in
the dynamics of positronium decays as expressed in the decay
amplitude. It is clear that the physical phase space for these decays is
determined by the masses of external particles, in particular the initial
available energy $M$. Thus, we will study the effects of these
non-perturbative corrections in observables associated to positronium
decays, which contain the BE effects in the dynamics (amplitude) and the
kinematics (phase-space).

\begin{center}
\bf 2$\gamma$ decay of parapositronium.
\end{center}

The decay amplitude for p-Ps$ \rightarrow \gamma \gamma$ obtained from Eq.
(16) can be expressed as:
\begin{equation}
\overline{\cal M}^{\mu\nu} (B(^1S) \rightarrow \gamma \gamma) =
2Ce^2\Phi_0 \frac{\varepsilon^{\mu\nu\alpha\beta}
l_{1\alpha}P_{\beta}}{P\cdot l_1}{\cal A}_1\ .
\end{equation}
The corresponding rate for this two-body decay where photons fly
apart with energy $M/2$ in the rest frame of p-Ps, is given by
($2P.l_1=M^2$):
\begin{eqnarray}
\Gamma(\mbox{\rm p-Ps} \rightarrow \gamma\gamma) &=& \frac{1}{2} \alpha^5m
\times
\frac{4m^2}{M^2} \left[ \frac{1}{1+2\gamma^2/M^2} \right]^2 \nonumber \\
&\approx& \frac{1}{2} \alpha^5m \left(1- \frac{\alpha^4}{64} \right),
\end{eqnarray}
where we have used the definition of $\gamma^2$ as given after Eq. (9) 
 and we have
neglected corrections of $O(\alpha^6)$. Thus, (non-perturbative) BE
  corrections for $2\gamma$ decays of parapositronium are negligible
small and are beyond present experimental precision.

\begin{center}
\bf para-dimuonium (p-Dm) decay into $e^+e^-\gamma$
\end{center}

  The decay mechanism for this process is similar to the previous one,
but where one virtual photon converts into a $e^+e^-$ pair. Therefore the
corresponding amplitude is:
\begin{equation}
\overline{\cal M}^{\mu} (\mbox{\rm p-Dm} \rightarrow e^+e^-\gamma ) =
2Ce^3\Phi_0 \frac{\varepsilon^{\mu\nu\alpha\beta}
l_{1\alpha}P_{\beta}}{P\cdot l_1}{\cal A}_1
\frac{\left\{ \bar u(p) \gamma_{\nu} v(p')\right\} }{r^2},
\end{equation}
where $l_1$ and $r\equiv p+p'=P-l_1$ denote, respectively, the
four-momenta of the photon and the $e^+e^-$ pair. 

   The single photon spectrum in this case is given by
($a\equiv4m_e^2/M^2,\ x=2E_{\gamma}/M$):
\begin{equation}
\frac{d\Gamma(\mbox{\rm p-Dm} \rightarrow  e^+e^-\gamma )}{dx} =
\frac{16\alpha^3\Phi_0^2}{3M^2} \sqrt{1-\frac{a}{1-x}} \left[ a+2(1-x)
\right] \frac{x^3}{(1-x)^2} \left(x+\frac{2\gamma^2}{M^2} \right)^{-2} \ .
\end{equation}
Observe that this spectrum falls as $x^3$ when $x\rightarrow 0$ due to
non-zero BE effects, instead of the usual behavior
(proportional to $x$) expected when these effects are neglected. It is
interesting to note that when $x\ll 2\gamma^2/M^2$, Eq. (21) behaves as
the corresponding photon energy spectrum in $\pi^0 \rightarrow
e^+e^-\gamma$ decay, for a point-like pion vertex. Thus, BE 
corrections affect the shape of the spectrum or, conversely, actually
probes the structure of the bound state.  Indeed, Eq. (21) can be seen as
a result that extrapolates the spectrum between the $e^+e^-\gamma$ decay
of a pseudoscalar point particle ($\pi^0$ case) and the standard bound
state calculations (paradimuonium decay without BE corrections).

  A closed (but long) analytic expression for the decay rate can be
obtained from integration of Eq. (21). A useful approximation that takes
into account leading non-vanishing corrections of $O(\alpha^2)$ is:
\begin{equation}
\Gamma(\mbox{\rm p-Dm} \rightarrow e^+e^-\gamma) = \frac{\alpha^6m}{6\pi}
\left[
F_0-(1-a)^{3/2}\frac{\alpha^2}{2} +O(\alpha^4) \right]\ ,
\end{equation}
where the function $F_0 \equiv (4/3)\sqrt{1-a}(a-4)+2\ln
[(1+\sqrt{1-a})/(1-\sqrt{1-a})]$ fixes the lowest order rate.

\begin{center}
\bf 3$\gamma$ decays of orthopositronium
\end{center}

  The squared unpolarized amplitude for o-Ps$ \rightarrow
\gamma\gamma\gamma$ obtained from Eq. (17), in terms of dimensionless
photon energy variables $x_i=2E_{\gamma i}/M$ ($x_1+x_2+x_3=2$) is given
by:
\begin{equation}
\sum_{pols}|{\cal M}(3\gamma)|^2 \propto 
\left[ \left(\frac{1-x_1}{x_2x_3} \right)^2+\left(\frac{1-x_2}{x_1x_3}
\right)^2+\left(\frac{1-x_3}{x_1x_2} \right)^2 \right] \prod_{i=1}^3 
\left( \frac{x_i}{x_i+\frac{2\gamma^2}{M^2}} \right)^2  .
\end{equation}
As in the previous case, the single photon spectrum will be softened by 
BE corrections due to the last factor in the squared amplitude.

  The phase space integration of Eq. (23) can be performed in analytic
form and expressed in terms of dilogarithmic functions. However, it is
more illustrative to express the decay rate in terms of an expansion in
powers of $\alpha^2$ (or $\gamma^2$). If we express the decay rate in
terms of the mass $m$ of constituents, we obtain:
\begin{equation}
\Gamma(\mbox{\rm o-Ps} \rightarrow 3\gamma) \approx \alpha^6 m
\frac{2(\pi^2-9)}{9\pi} \left[ 1-\frac{5}{4}\alpha^2 \right]\ .
\end{equation}
Thus, BE corrections affects the decay rate of
orthopositronium at order $\alpha^2$ which are indeed relevant when
confronted to accuracy of present experiments. A comparison of Eqs. (4)
and (24) indicates that the net effects of BE corrections is to
resize the coefficient appearing in nonlogarithmic $O(\alpha^2)$
corrections of orthopositronium, namely:
\begin{equation}
B_0 \rightarrow B_0-\frac{5\pi^2}{4}\approx 44.52(26)-12.34 \ .
\end{equation}
{\it i.e.} a non-negligible 28\% correction at order $\alpha^2$.

Before concluding, let us address some comments on possible implications
of BE corrections in analogous decays of quarkonia. Notice that BE in
quarkonia is not simply related to the masses  of the constituent quarks
and to the coupling $\alpha_s$ as in Eq. (8) due to confinement.
If we assume, however, that BE corrections do affect
quarkonium decays in a similar form as in positronium, we can address some
apparent conflicts in some of their inclusive hadronic and radiative
decays. First, the 
photon spectrum measured in $J/\psi \rightarrow gg\gamma$ decays seems to
be softer than predicted by perturbative QCD \cite{soft} which indicate
possible large non-perturbative effects. This is precisely the effect
induced by BE corrections in the single photon spectrum of
o-Ps$\rightarrow  \gamma\gamma\gamma$. Second, different inclusive rates
in quarkonium decays as discussed in \cite{lowalpha} will, in general, get
decreased by BE corrections. This may increase the
relatively low values of the strong coupling constants extracted from
 ratios of experimental quarkonium rates \cite{lowalpha}.  Finally, these
effects will also manifest in the ``14 \% rule" observed in the ratio
$BR(\psi(2S) \rightarrow X)/BR(J/\psi \rightarrow X)$ for single 
photon mediated decays \cite{rule14}, because BE are different for these
two radial excitations of charmonium.

   In this paper we have computed binding energy corrections to
positronium decays in a specific ansatz where these non-perturbative
effects are introduced as a necessity to account for analyticity and gauge
invariance of the corresponding QED amplitudes. The BE corrections (of
order $\alpha^2$) to orthopositronium decay are indeed relevant in view of
recent efforts to achieve a precise comparison of theory and experiment.
When extended to the quarkonium sector, these BE corrections may
contribute \cite{next} to solve apparent discrepancies observed between
experimental data and perturbative QCD calculations of some inclusive
rates of quarkonia.

{\large Acknowledgements}: C.S. acknowledges financial support from
FNRS (Belgium). G.L.C. was partially supported by Conacyt (M\'exico)
under contract No. 32429.

\newpage
\begin{figure}
\label{Figure 1}
\centerline{\epsfig{file=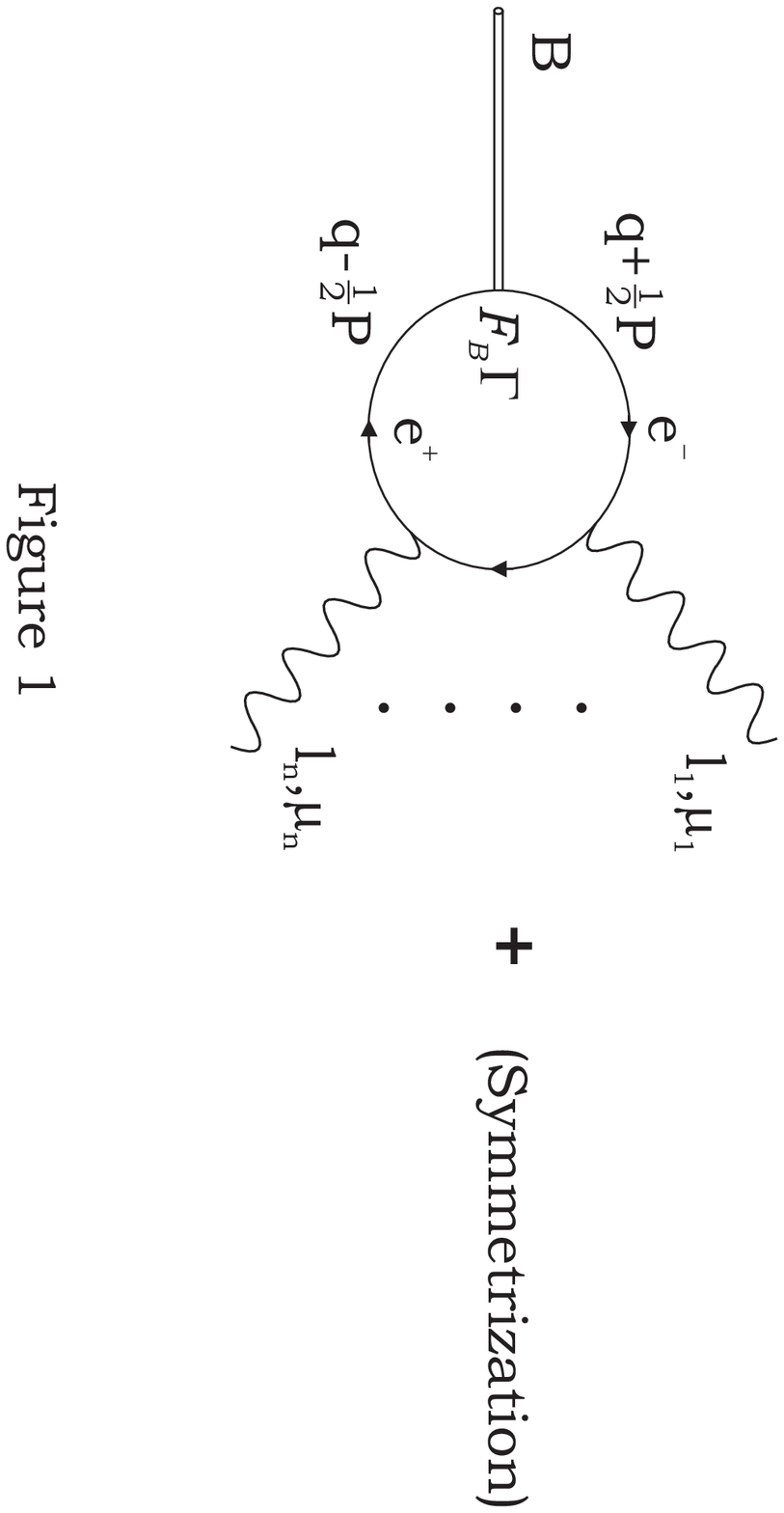,angle=0,width=6.5in}}
\vspace{-1.0in}
\end{figure}

\end{document}